\documentclass[letterpaper]{article}

\usepackage{natbib,lib/alifeconf}  %% The order is important
\usepackage{amsmath,amssymb}
\usepackage{url,hyperref,cleveref}
\usepackage{booktabs}

\usepackage{lipsum}

\usepackage{bm}
\usepackage{caption}
\usepackage{xcolor}
\usepackage{algorithm}
\usepackage{algorithmic}

\newcommand\blfootnote[1]{%
  \begingroup
  \renewcommand\thefootnote{}\footnote{#1}%
  \addtocounter{footnote}{-1}%
  \endgroup
}

\title{Social Reality Construction via Active Inference: Modeling the Dialectic of Conformity and Creativity}
\author{
    Kentaro Nomura$^{1,\dag}$,
    Takato Horii$^{1,2}$\\
    \mbox{}\\
    $^1$The University of Osaka, Japan, 
    $^2$IRCN, The University of Tokyo, Japan \\
    $\dag$nomura.kentaro.7ks@ecs.osaka-u.ac.jp
} % email of corresponding author

\begin{document}

\maketitle

\begin{abstract}
% Abstract length should not exceed 250 words
Social agents both internalize collective norms and reshape them through creative action, yet computational models have not captured this bidirectional process within a unified framework.
We propose a multi-agent simulation model grounded in active inference that formalizes the dialectical constitution of social reality on a structured social network.
Each agent maintains an internal generative model, communicates with neighbors to form social priors, creates novel observations, and selectively incorporates others' creations into memory.
Simulation experiments demonstrate three main findings.
First, informationally cohesive social groups emerge endogenously, with representational alignment mirroring the cluster topology of the underlying network.
Second, a circular mutual constitution arises between social representations and the observation distribution, maintained through agents' creative acts that project representational structure onto the external world.
Third, the propagation of creations exhibits selective, heterogeneous patterns distinct from the stable diffusion of social representations, indicating that agents construct cultural niches through local interaction dynamics.
These results suggest that the interplay between social conformity and creative deviation can give rise to the endogenous formation and differentiation of shared social reality.
\end{abstract}

% Choose one of: Full Paper, Summaries, or Late Breaking Abstracts 
% Submission type: \textbf{Full Paper}\\

% If sharing code / data, anonymize your repository and paste the link here.
% Example of anonymizing sevice for github: https://anonymous.4open.science/
% delete this line if not needed
Data/Code available at: \url{https://github.com/jemand-rkn/social-reality-aif.git}
\blfootnote{\textcopyright  2026 Kentaro Nomura, Takato Horii. Published under a Creative Commons Attribution 4.0 International (CC BY 4.0) license.}

\section{Introduction}

Social agents both internalize collective norms and reshape them by creating and reifying new objects of meaning in the physical world.
Through local interaction, these individual acts give rise to a shared social reality that, in turn, feeds back to shape further individual action.
% Berger and Luckmann \citep{Berger1966-nm} described this process as a dialectic of externalization, objectivation, and internalization, while Mead \citep{mead1972mind} formulated the same tension as the interplay between the norm-internalizing \textit{Me} and the creatively responding \textit{I}.
\citet{mead1972mind} located this tension within the individual, formulating it as the interaction between the norm-internalizing \textit{Me} and the creatively responding \textit{I}.
\citet{Berger1966-nm}, in contrast, characterized the same dynamic at the macro level as a dialectic of externalization, objectivation, and internalization, describing how society reproduces itself through individual action.
% Despite this rich theoretical tradition, the computational mechanisms by which top-down social regulation and bottom-up creative formation jointly produce stable yet evolving norms remain poorly understood.
Despite this rich theoretical tradition, the computational mechanisms by which bottom-up creative formation and top-down social regulation jointly produce stable yet evolving norms remain poorly understood.
To address this gap, we propose a multi-agent simulation model grounded in the free energy principle.
% In this model, agents communicate via a naming game on a social network and actively create novel observations, thereby computationally realizing the dialectical process of social reality construction.
In this model, agents actively create novel observations and communicate via a naming game on a social network, thereby computationally realizing the dialectical process of social reality construction.

Existing models in the Artificial Life community have demonstrated that shared conventions can emerge from local agent interactions, yet they uniformly treat agents as rule-following entities rather than probabilistic reasoners with internal generative models.
The naming game paradigm \citep{Steels1995-tw, Baronchelli2006-lx} established that shared vocabularies self-organize through pairwise interaction, undergoing a sharp phase transition to global consensus without centralized coordination.
Similarly, the iterated learning framework \citep{Smith2003-my, Kirby2008-zb} showed that cultural transmission alone can generate compositional linguistic structure.
Communicative exploration has also been shown to sustain open-ended collective novelty \citep{Witkowski2019-hr}.
However, because agents in these models lack internal generative models, they cannot capture the top-down regulatory influence of social context on individual cognition or the generative capacity to produce observations that deviate from prevailing norms.

Active inference \citep{Parr-aif} offers a principled framework for deriving collective behavior and social norm formation from individual surprise minimization without requiring explicit behavioral rules.
Canonical collective phenomena---cohesion, milling, and directed motion---have been shown to emerge when agents minimize surprise about their conspecifics, with real-time model updates coupling individual learning to group-level properties \citep{Heins2024-fm}.
Communication under active inference has further been shown to drive generalized synchrony and long-term convergence of generative models between agents \citep{Friston2015-jr, Friston2015-fp}.
Norm acquisition has also been formalized as free energy minimization with respect to a model of others' expectations \citep{Veissiere2019-jy, Kaufmann2021-wy}.
However, these models address either spatial coordination or the internalization of pre-existing expectations; none accounts for how agents' own creative acts feed back into the formation of shared social representations.
This gap corresponds to the missing bottom-up direction in niche construction theory \citep{Odling-Smee2003-ta, Laland2000-fu, Laland2011-et}, which holds that organisms actively reshape their environment.
This process has been formalized as free energy minimization through environmental modification, providing a variational basis for niche construction \citep{Constant2018-zc}.

The Collective Predictive Coding (CPC) hypothesis \citep{Taniguchi2024-ym} extends predictive coding and the free energy principle to the societal scale, offering a theoretical account of how shared representations such as symbol systems and social norms emerge through communicative interaction.
Under CPC, social representations are formed through decentralized Bayesian inference.
The Metropolis--Hastings Naming Game (MHNG) has been proposed as its concrete algorithmic realization \citep{Hagiwara2019-cm, Taniguchi2023-lj}, which has been further extended to $N$-agent populations through a recursive formulation \citep{Inukai2023-fq}.
However, CPC treats social representations as external shared variables that connect all agents through a joint generative model.
This formulation implicitly assumes that a common representation is shared across the population, thereby precluding the natural emergence of distinct social groups.
Furthermore, agents in this framework are passive observers that infer representations from observations they \textit{receive}, leaving the feedback pathway from individual creative acts to social representation formation unaddressed.
In the present work, social representations are instead placed inside each agent and allowed to differ across individuals, enabling representational alignment to emerge naturally from local interaction.
Agents are also endowed with the capacity for active creation, closing the feedback loop between individual creativity and collective norm formation.

% We present a multi-agent simulation model that formalizes the bidirectional constitution of social reality within a unified active inference framework on a structured social network.
% Each agent maintains an internal generative model, communicates with neighbors to acquire knowledge of the shared social context, actively creates novel observations, and selectively incorporates others' creations into its memory.
% Top-down regulation and bottom-up formation are realized as variational and expected free energy minimization, respectively, placing them in a formally adversarial yet ultimately complementary relationship: social conformity provides a stable representational scaffold from which creative exploration departs, while creative action diversifies the observational landscape.
% Simulation experiments show that this productive tension gives rise to the endogenous formation of informationally cohesive social groups, a circular mutual constitution of social representations and the observation distribution, and the selective construction of cultural niches through heterogeneous propagation of creations across the network.
We present a multi-agent simulation model that formalizes the bidirectional constitution of social reality within a unified active inference framework on a structured social network.
Each agent is formalized as an internal generative model that is continuously updated through communication with neighbors, active creation of novel observations, and selective incorporation of others' creations.
Top-down regulation and bottom-up formation are realized as variational and expected free energy minimization, respectively, placing them in a formally adversarial yet complementary relationship.
The present work makes three contributions.
First, we show that informationally cohesive social groups emerge endogenously, mirroring the cluster topology of the underlying network.
Second, social representations and the observation distribution are mutually constituted through agents' creative acts.
Third, creations propagate in selective, heterogeneous patterns, indicating that agents construct cultural niches through local interaction.

\section{Proposed Model}

\subsection{Agent Architecture}\label{subsec:agent-architecture}

Each agent in the proposed model maintains an internal generative model that describes how observations are generated from social representations as latent variables.
Here, observations represent information that agents experience in the physical world.
The generative model is agent-specific and is updated in a fully decentralized manner.
Specifically, the generative model for agent $A^k$ is defined as follows:
\begin{align}
    \text{social prior:~} &\bm{z} \sim p^k_{\textrm{social}}(\bm{z}),\\
    \text{likelihood:~} &\bm{o} \sim p^k_\theta(\bm{o}|\bm{z}),\\
    \text{posterior:~} &\bm{z} \sim q^k_\phi(\bm{z}|\bm{o}),
\end{align}
where $\bm{z}$ and $\bm{o}$ denote the social representation and the observation, respectively, and $k$ indexes the agent.
The likelihood $p^k_\theta(\bm{o}|\bm{z})$ and the approximate posterior $q^k_\phi(\bm{z}|\bm{o})$ are parameterized by neural network weights $\theta$ and $\phi$, respectively.
The social prior $p^k_{\textrm{social}}(\cdot)$ is a probability distribution that reflects the social context in which the agent is situated.
This distribution evolves as the agent interacts with others.

Each agent also possesses a discriminator $D^k_\psi(\bm{z}, \bm{o})$ that quantifies the divergence between the agent's approximate posterior $q^k_\phi(\cdot|\bm{o})$ and the social prior $p^k_{\textrm{social}}(\cdot)$.
The discriminator is trained to approximate the log-density ratio between these two distributions:
\begin{equation}
    D^k_\psi(\bm{z}, \bm{o}) \approx \log\frac{q^k_\phi(\bm{z}|\bm{o})}{p^k_{\textrm{social}}(\bm{z})}.
\end{equation}
The discriminator enables each agent to assess how much its own observation-conditioned beliefs deviate from the social expectations formed through interaction with its neighbors.
This discriminator plays a central role in both model updating and the creation of novel observations.

% Each agent further maintains a memory buffer $\mathcal{B}^k$ that stores observations as a first-in-first-out (FIFO) data structure.
% Newly created observations are appended to the buffer, while the oldest entries are discarded.
% In addition to the agent's own creations, observations produced by other agents may be incorporated into $\mathcal{B}^k$ through selective replacement.
Each agent further maintains a memory buffer $\mathcal{B}^k$ that stores artifacts as a first-in-first-out (FIFO) data structure.
Newly created artifacts are appended to the buffer, while the oldest entries are discarded.
In addition to the agent's own creations, artifacts produced by other agents may be incorporated into $\mathcal{B}^k$ through selective replacement.
When the agent draws samples from $\mathcal{B}^k$ for model updating or communication, the stored artifacts serve as observations for its generative model.

% \subsection{Creation of Novel Observations via Active Inference}\label{subsec:creation}
\subsection{Creation of Novel Artifacts via Active Inference}\label{subsec:creation}

% Each agent creates novel observations by sampling an observation $\bm{o}_+$ that minimizes the expected free energy within the framework of active inference.
Each agent creates novel artifacts $\bm{o}_+$ that minimize the expected free energy within the framework of active inference.
% Note that, we formulate this process as directly sampling observations that minimize the expected free energy, without introducing explicit actions, thereby simplifying the model.
The expected free energy $\mathcal{G}^k$ for agent $A^k$ is defined as:
\begin{equation}\label{eq:efe}
\begin{aligned}
    \mathcal{G}^k(\bm{o}_+) &= -D_{\mathrm{KL}}[q^k_\phi(\bm{z}|\bm{o}_+)\|p^k_{\textrm{social}}(\bm{z})] - \mathbb{E}_{q^k_\phi}[\log p^k_\theta(\bm{o}_+|\bm{z})]\\
    &\approx -\mathbb{E}_{q^k_\phi}[D^k_\psi(\bm{z}, \bm{o}_+)] - \mathbb{E}_{q^k_\phi}[\log p^k_\theta(\bm{o}_+|\bm{z})].
\end{aligned}
\end{equation}
In practice, the second term is weighted by a coefficient $\lambda$.
The optimization is performed via gradient descent, initialized from an artifact sampled from the agent's memory $\mathcal{B}^k$, and the resulting artifact is subsequently appended to $\mathcal{B}^k$.

The two terms in Eq.~\eqref{eq:efe} jointly drive agents to explore artifacts that are novel relative to the prevailing social context while remaining consistent with their own generative models.
The first term evaluates the negative divergence between the approximate posterior conditioned on the created artifact $\bm{o}_+$ and the social prior.
Minimizing this term encourages the agent to produce creations that deviate from the established social prior, thereby maximizing epistemic information gain regarding social representations.
The second term corresponds to the reconstruction error of $\bm{o}_+$ under the agent's generative model, which penalizes excessive deviation from the agent's individual knowledge.
Through simultaneous optimization of both terms, each agent produces artifacts that represent physical entities not yet captured by the prevailing social context, while preserving the epistemological consistency of its own model.
Note that we formulate this process as directly sampling observations that minimize the expected free energy, without introducing explicit actions, thereby simplifying the model.

\subsection{Model Update}\label{subsec:model-update}

The discriminator $D^k_\psi(\bm{z}, \bm{o})$ is trained under the $f$-GAN framework~\citep{Nowozin2016-bi}, which generalizes generative adversarial networks~\citep{Goodfellow2014-ka} to $f$-divergences for implicitly learning unknown probability distributions through adversarial training.
Specifically, we adopt the reverse Kullback--Leibler divergence as the $f$-divergence to estimate the divergence between the social prior and the approximate posterior.
Let $\mathcal{P}^k$ denote the set of observation--representation pairs sampled from the social prior through inter-agent communication.
The loss function for training $D^k_\psi$ is:
\begin{equation}
\begin{split}
    \mathcal{L}^k(\psi) = &\mathbb{E}_{(\bm{o},\bm{z})\sim \mathcal{P}^k}[\exp(D^k_\psi(\bm{z}, \bm{o}))] \\
    &- \mathbb{E}_{\bm{o}\sim\mathcal{B}^k,\, q^k_\phi(\bm{z}|\bm{o})}[D^k_\psi(\bm{z}, \bm{o})].
\end{split}
\end{equation}

The generative model parameters $\theta$ and $\phi$ are updated by minimizing the variational free energy $\mathcal{F}^k$, which is expressed using the discriminator as:
\begin{equation}\label{eq:vfe}
    \begin{aligned}
        \mathcal{F}^k(\theta, \phi) &= D_{\mathrm{KL}}[q^k_\phi(\bm{z}|\bm{o})\|p^k_{\textrm{social}}(\bm{z})] - \mathbb{E}_{q^k_\phi}[\log p^k_\theta(\bm{o}|\bm{z})]\\
        &\approx \mathbb{E}_{q^k_\phi}[D^k_\psi(\bm{z}, \bm{o})] - \mathbb{E}_{q^k_\phi}[\log p^k_\theta(\bm{o}|\bm{z})].
    \end{aligned}
\end{equation}
In practice, the first term is weighted by a coefficient $\beta$.
Because Eq.~\eqref{eq:vfe} contains the positive value of $D^k_\psi$ in the first term, minimizing the variational free energy drives each agent's approximate posterior toward the social prior, thereby aligning the agent's internal model with the prevailing social context.

The model update and the creation process stand in an adversarial relationship.
Comparing Eq.~\eqref{eq:vfe} with Eq.~\eqref{eq:efe} reveals that the sign of the term involving $D^k_\psi$ is reversed between the two objectives.
Variational free energy minimization during model updating pushes the approximate posterior toward the social prior, whereas expected free energy minimization during creation and selective memorization pushes it away.
The former thus constitutes a top-down regulatory influence from the social context, while the latter represents a bottom-up contribution to social formation through exploratory creative acts.

\subsection{Social Network}\label{subsec:social-network}

The agents described above are situated on a social network that governs their communication topology and determines the scope of social interaction.
The network is formalized as an undirected graph $(\mathcal{K}, \mathcal{E})$, where $\mathcal{K} = \{A^1, \ldots, A^K\}$ denotes the set of $K$ agents and $\mathcal{E}$ denotes the set of edges.
At each time step, agent $A^k$ exchanges inferred social representations and creations with its neighboring agents $A^{k'} \in \mathcal{N}(k)$, where $\mathcal{N}(k)$ is the set of agents connected to $A^k$ by an edge.
Through these pairwise exchanges, each agent's social prior and memory are shaped not only by its own internal processes but also by the creations and representations of its neighbors, as described in the following subsections.

\subsection{Selective Memorization of Others' Creations}\label{subsec:incorporation}

Whether to accept a creation received from another agent and update one's memory can be understood as action selection under active inference.
In active inference, an agent selects actions $a$ according to a Boltzmann distribution over the expected free energy: $P(a) \propto \exp(-\mathcal{G}^k(a)/\tau)$, where $\tau$ is a temperature parameter controlling the stochasticity of the selection~\citep{Parr-aif}.
We consider two candidate actions for each received creation: acceptance ($a = \bm{o}'_+$) and rejection ($a = \bm{o}_-$).
Here, $\bm{o}'_+$ denotes a creation received from a neighboring agent in $\mathcal{N}(k)$, and $\bm{o}_- \sim \mathcal{B}^k$ is an observation sampled from the agent's own memory.
Evaluating $\mathcal{G}^k$ for both candidates and comparing their energy values yields the acceptance probability:
\begin{equation}
    r_o = \min\left(1, \exp\left(-\frac{\mathcal{G}^k(\bm{o}'_+) - \mathcal{G}^k(\bm{o}_-)}{\tau}\right)\right),
\end{equation}
and upon acceptance, the memory buffer is updated as:
\begin{equation}
    \mathcal{B}^k \gets (\mathcal{B}^k \setminus \{\bm{o}_-\}) \cup \{\bm{o}'_+\}.
\end{equation}
In the limit $\tau \to 0$, the agent deterministically selects the creation with lower expected free energy.
This pairwise comparison enables each agent to judge whether another agent's creation better aligns with its current interests than its own past creations, and to selectively retain the more relevant artifact in memory.

\subsection{Communication via Social Representations}\label{subsec:communication}

Agents construct their social priors by exchanging social representations with neighbors on the social network and integrating the received representations with their own models.
Through this communicative process, each agent forms prior knowledge that reflects the perspective of a generalized social member as perceived from its own viewpoint.
% We assume joint attention during communication: both communicating agents present their memorized observations and creations, and infer social representations conditioned on these shared stimuli.
We assume joint attention during communication: both communicating agents present their memorized artifacts and creations, and infer social representations conditioned on these shared artifacts.

To obtain samples from each agent's social prior $p^k_{\textrm{social}}(\bm{z})$ under joint attention, we employ the Metropolis--Hastings Naming Game (MHNG)~\citep{Taniguchi2023-lj}.
% MHNG is a method for sampling latent variables that can be shared between two agents who jointly attend to the same observations.
MHNG is a method for sampling latent variables that can be shared between two agents who jointly attend to the same artifacts.
Although MHNG originally yields samples from an approximate posterior $q^k(\bm{z})$, we identify these with the social prior via Bayesian updating, i.e., $p^k_{\textrm{social}}(\bm{z}) = q^k(\bm{z})$.
Let $\mathcal{O}^{k,k'} = \mathcal{C}^k \cup \mathcal{D}^k \cup \mathcal{C}^{k'} \cup \mathcal{D}^{k'}$ denote the union of creations $\mathcal{C}^\star$ and memory samples $\mathcal{D}^\star \subset \mathcal{B}^\star$ from agent $A^k$ and its neighbor $A^{k'} \in \mathcal{N}(k)$.
Notably, not only the creations $\mathcal{C}^\star$ but also past memories $\mathcal{D}^\star$ serve as objects of joint attention.

Both agents infer social representations $\bm{z}$ from observations in $\mathcal{O}^{k,k'}$ using their respective inference models $q^\star_\phi$.
Let $\bm{z}^\star \sim q^\star_\phi(\cdot|\bm{o})$ denote the social representation inferred by agent $A^\star$ from $\bm{o} \sim \mathcal{O}^{k,k'}$.
Agent $A^k$ accepts the representation $\bm{z}^{k'}$ received from its neighbor with probability:
\begin{equation}
    r_z = \min\left(1, \frac{p^k_\theta(\bm{o}|\bm{z}^{k'})}{p^k_\theta(\bm{o}|\bm{z}^k)}\right),
\end{equation}
and upon rejection retains its own inferred representation $\bm{z}^k$ as a sample from $p^k_{\textrm{social}}(\bm{z})$.

The set of observation--representation pairs $\mathcal{P}^k$ used for discriminator training is constructed through this communication process.
Each pair in $\mathcal{P}^k$ consists of an observation drawn from $\bigcup_{k'=\sharp(A\in\mathcal{N}(k))} \mathcal{O}^{k,k'}$ and the corresponding social representation accepted or retained through the Metropolis--Hastings procedure above.

\subsection{Simulation Procedure}\label{subsec:simulation-flow}

\begin{algorithm}[t]
\caption{Simulation procedure of the proposed model}
\label{alg:simulation}
\begin{algorithmic}[1]
\FOR{each time step $t = 1, 2, \ldots$}
    \STATE \textbf{// Creation}
    \FOR{each agent $A^k \in \mathcal{K}$}
        \STATE Create $\bm{o}_+ \gets \arg\min_{\bm{o}} \mathcal{G}^k(\bm{o})$; append to $\mathcal{B}^k$
    \ENDFOR
    \STATE \textbf{// Selective memorization}
    \FOR{each agent $A^k \in \mathcal{K}$}
        \STATE Receive $\bm{o}'_+$ from $A^{k'} \in \mathcal{N}(k)$; accept into $\mathcal{B}^k$ with prob.\ $r_o$
    \ENDFOR
    \STATE \textbf{// Communication via MHNG}
    \FOR{each edge $(A^k, A^{k'}) \in \mathcal{E}$}
        \STATE Exchange representations $\bm{z}^k, \bm{z}^{k'}$ over joint observations $\mathcal{O}^{k,k'}$
        \STATE Accept $\bm{z}^{k'}$ with prob.\ $r_z$; collect pairs into $\mathcal{P}^k$
    \ENDFOR
    \STATE \textbf{// Model update}
    \FOR{each agent $A^k \in \mathcal{K}$}
        \STATE Update $\psi^k$ by minimizing $\mathcal{L}^k(\psi)$; update $\theta^k, \phi^k$ by minimizing $\mathcal{F}^k(\theta, \phi)$
    \ENDFOR
\ENDFOR
\end{algorithmic}
\end{algorithm}

The proposed model iterates through four stages at each time step, cycling between creative exploration and social conformity.
% Algorithm~\ref{alg:simulation} summarizes the complete procedure.
Algorithm~\ref{alg:simulation} summarizes the overall procedure.
First, each agent infers a novel artifact $\bm{o}_+$ by minimizing the expected free energy $\mathcal{G}^k$ and appends it to its memory $\mathcal{B}^k$.
Second, creations received from neighboring agents are selectively incorporated into $\mathcal{B}^k$ under active inference.
Third, each agent executes MHNG with its neighbors to obtain samples from the social prior $p^k_{\textrm{social}}(\bm{z})$.
Finally, the discriminator and generative model parameters $\theta$, $\phi$, and $\psi$ are updated.

Each stage fulfills a distinct functional role within the overall dynamics.
The creation and selective memorization stages modify the agent's memory $\mathcal{B}^k$.
The communication stage shapes the social prior.
The model update stage adapts the generative model and discriminator to the evolving social context.
Through the cyclic repetition of these stages, the formation of social representations and creative deviation mutually drive one another.

\section{Experiments}
We conducted a series of simulation experiments to verify that the proposed model gives rise to top-down regulatory effects from the collective society and bottom-up social formation through the creative acts of individual agents, and to observe what emergent phenomena result from these interactions.

\subsection{Experimental Settings}
\label{subsec:setting}
\begin{figure}
    \centering
    \includegraphics[width=\linewidth]{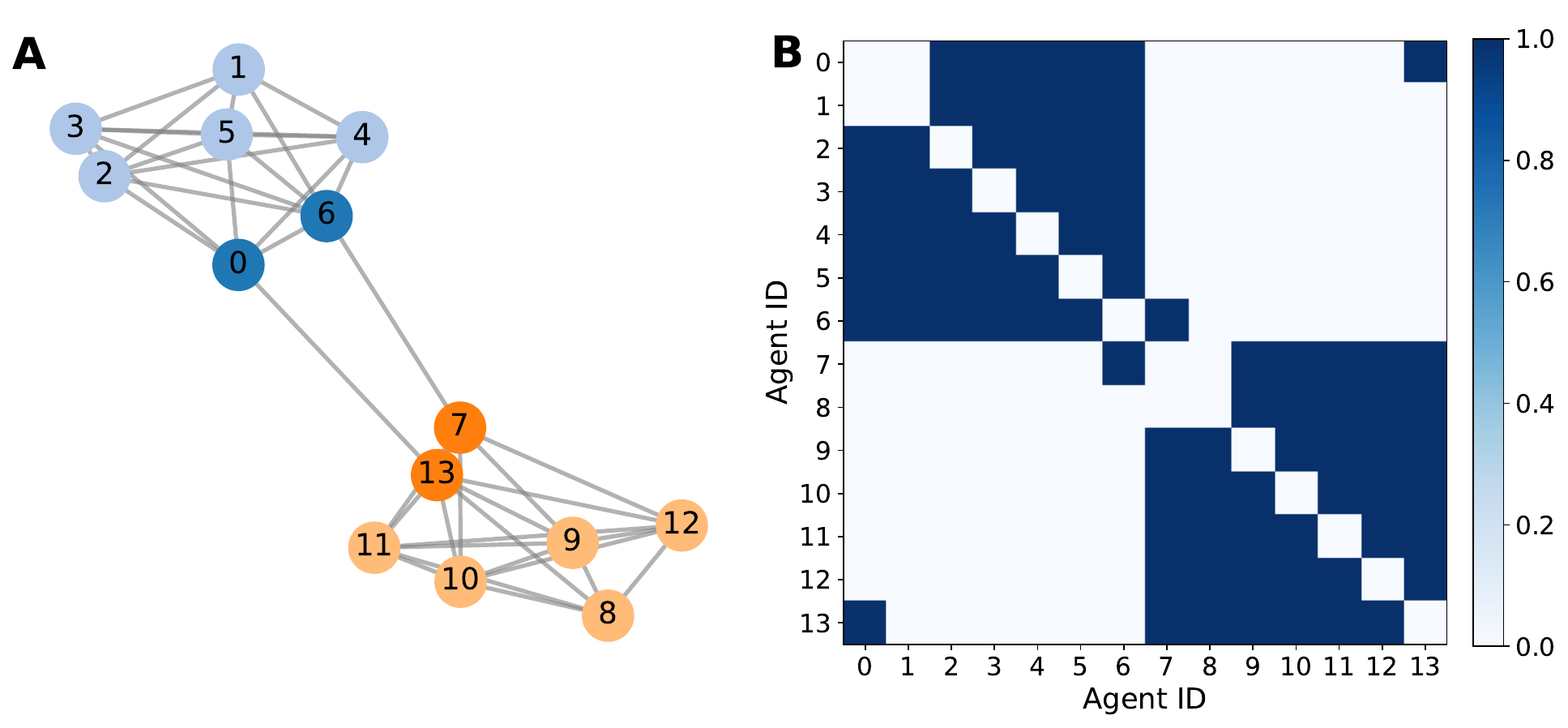}
    \caption{The social network used in the experiments.\\ \textbf{A}: Network structure (blue: cluster~0, orange: cluster~1).\\ \textbf{B}: Adjacency matrix.}
    \label{fig:social-network}
    \vspace{-4mm}
\end{figure}
The experiments were conducted on a connected caveman graph~\citep{Watts1999-pv} comprising $K = 14$ agents.
This social network topology consists of two cluster-structured subgraphs, each containing seven agents.
We refer to these subgraphs as cluster~0 (agents $k = 0$--$6$) and cluster~1 (agents $k = 7$--$13$).
Figure~\ref{fig:social-network} shows the network topology and its corresponding adjacency matrix.

Observations $\bm{o}$ were represented as two-dimensional continuous real-valued vectors, and social representations $\bm{z}$ as four-dimensional continuous real-valued vectors.
Each agent's memory buffer $\mathcal{B}^k$ had a capacity of 7{,}000 samples.
The buffer was initialized with samples drawn from Gaussian distributions whose peaks were placed at distinct locations in observation space for each agent (see the leftmost panel of Fig.~\ref{fig:memorized-obs}).
The generative model parameters $\theta$ and $\phi$ of each agent were initialized by pre-training on the initial memory $\mathcal{B}^k$ within the variational autoencoder~\citep{Kingma2014-vae} framework.
The discriminator parameters $\psi$ were initialized randomly.

The hyperparameters were set as follows.
The weighting coefficients for the second term of Eq.~\eqref{eq:efe} and the first term of Eq.~\eqref{eq:vfe} were set to $\lambda = 0.1$ and $\beta = 1.0$, respectively.
The temperature parameter of the Boltzmann distribution used in selective memorization was set to $\tau = 0.3$.

The simulation was run for 5{,}000 steps.
At each step, each agent created $|\mathcal{C}^k| = 6$ novel artifacts, and the number of samples drawn from memory $\mathcal{B}^k$ for communication was $|\mathcal{D}^k| = 100$.
Model parameters were updated for 5 iterations per step with a learning rate of $1 \times 10^{-5}$, using the Adam optimizer for both the creation and model-update phases.
The mini-batch size for sampling from $\mathcal{B}^k$ during model updates was set to 256.

\subsection{Experimental Conditions}\label{subsec:conditions}
% To assess the impact of the creation and selective memorization processes on the emergent social representations and their relationship to creative artifacts, we compared the following two experimental conditions:
To assess the impact of the creation and selective memorization processes on the emergent social representations and their relationship to creations, we compared the following two experimental conditions:
\begin{description}
    \item[\texttt{w/ creation}]
    Agents perform creation and selective memorization.
    The simulation follows the full procedure.
    \item[\texttt{w/o creation}]
    Agents perform neither creation nor selective memorization.
    During the communication phase, the set of creative artifacts shared in the MHNG is set to $\mathcal{C}^k = \emptyset$.
\end{description}

\section{Results}

\subsection{Temporal Evolution of Social Representations}
\begin{figure*}
    \centering
    \includegraphics[width=\linewidth]{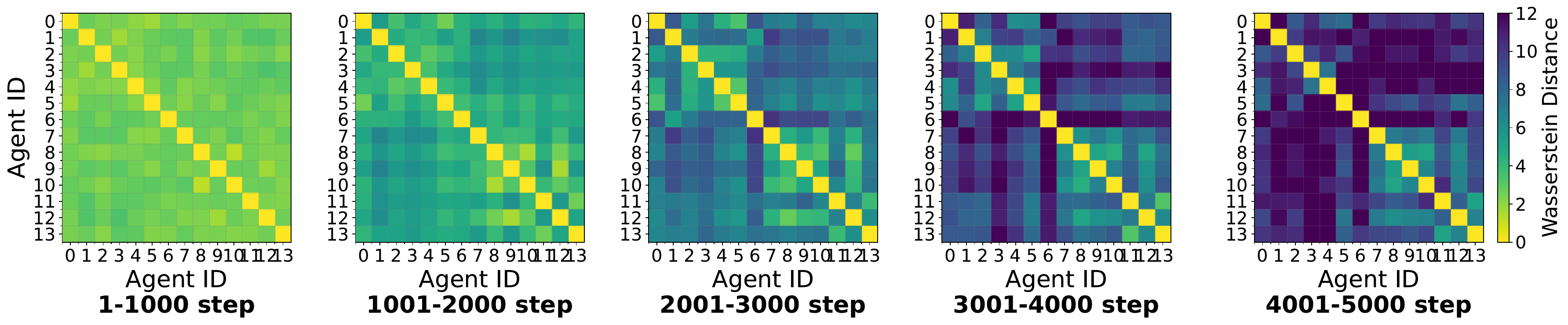}
    \caption{Evolution of the average Wasserstein distance matrix computed between the social representations inferred by agents.}
    \label{fig:wasserstein-snapshot}
    \vspace{-4mm}
\end{figure*}

% \begin{figure}[t]
%     \centering
%     \includegraphics[width=\linewidth]{fig/gw_mds_trajectory.pdf}
%     \caption{MDS embedding trajectory of the GW distance matrix computed from the inferred social representation structures. Numbers in black indicate time steps.}
%     \label{fig:gw-mds}
%     \vspace{-4mm}
% \end{figure}
\begin{figure}[t]
    \centering
    \includegraphics[width=\linewidth]{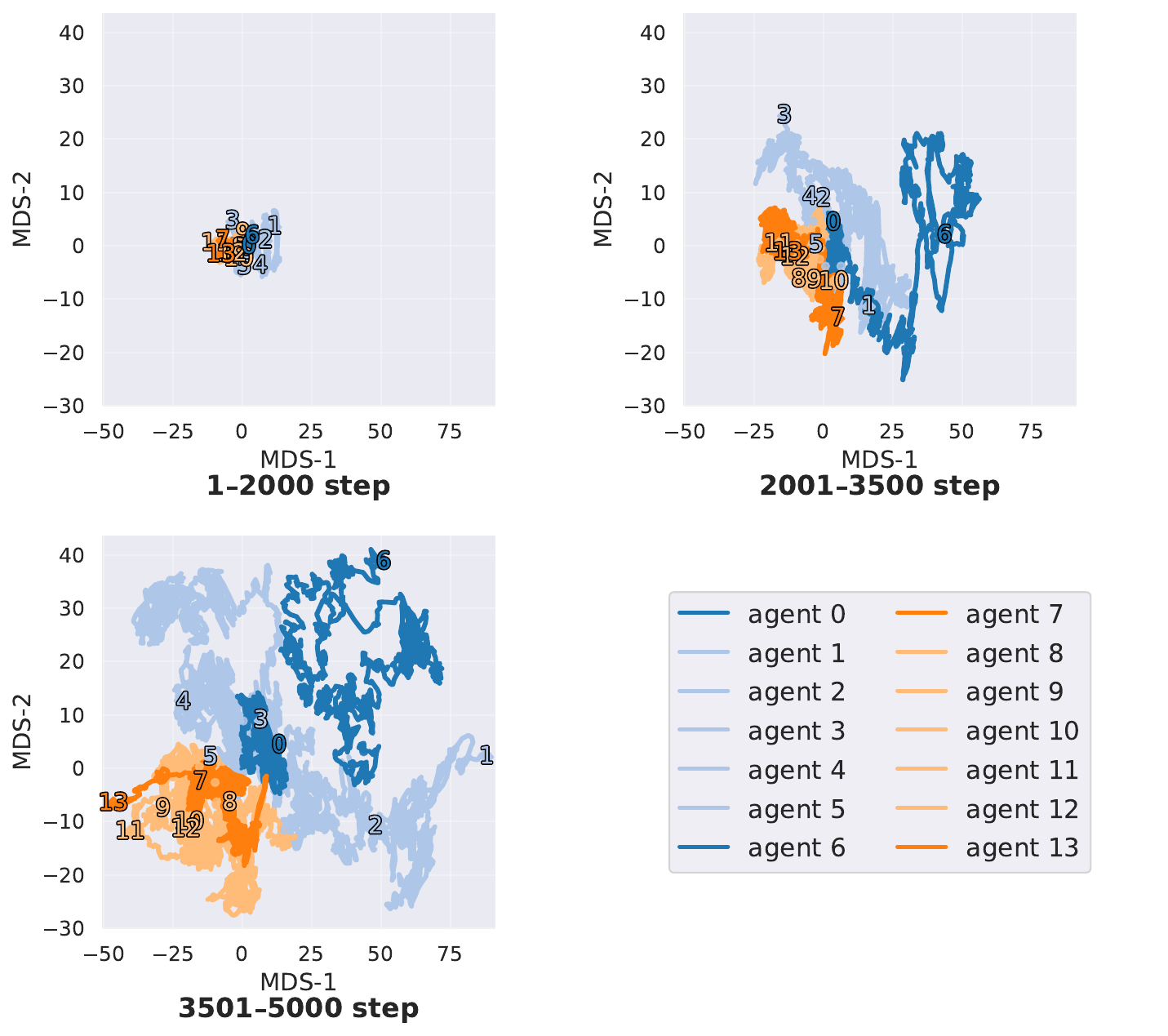}
    \caption{MDS embedding trajectory of the GW distance matrix from the inferred social representation structures.}
    \label{fig:gw-mds}
    \vspace{-4mm}
\end{figure}

We characterized the relational structure of social representations by comparing the representations inferred by each agent in the \texttt{w/ creation} condition.
We first constructed a reference observation set $\mathcal{O}_{\mathrm{ref}}$ by uniformly sampling observations from each agent's memory at each time step.
% Each agent's inference model was then applied to every element of $\mathcal{O}_{\mathrm{ref}}$, yielding the posterior distribution $q^k_\phi(\cdot|\bm{o})$ over social representations for all agents $k$.
Each agent's inference model was then applied to every element of $\mathcal{O}_{\mathrm{ref}}$, yielding the posterior distribution $q^k_\phi(\cdot|\bm{o})$ over social representations for all agents.
For every agent pair $(k, k')$, we computed the mean Wasserstein distance between the per-observation posteriors as a measure of how similarly two agents represent the same observation.

The Wasserstein distance matrix reveals a gradual transition from uniform inter-agent distances to a block-diagonal structure (Fig.~\ref{fig:wasserstein-snapshot}).
Initially, representational distances were roughly uniform across all agent pairs.
As the simulation progressed, intra-cluster distances decreased relative to inter-cluster distances, indicating that representational alignment mirrored the cluster structure of the social network.
As a result, informationally cohesive social groups---composed of frequently communicating individuals---emerged endogenously from the dynamics of the model.

However, Wasserstein distance captures only point-wise dissimilarity and does not reflect whether the relational geometry among representations is shared across agents, even when absolute values differ.
To evaluate this structural similarity, we additionally computed the Gromov--Wasserstein (GW) distance \citep{Memoli2011-kn}, which quantifies the degree to which the relational structure of an entire set of representations is preserved across agents.

The GW distance, visualized through multidimensional scaling (MDS) embedding, reveals a systematic, cluster-aligned divergence in representational structure over time (Fig.~\ref{fig:gw-mds}).
Initially, all agents were densely clustered near the origin, indicating a shared relational structure.
Over the course of the simulation, agents in cluster~0 migrated toward the upper region of the embedding space, while those in cluster~1 moved downward, reflecting a divergence that mirrors the community boundaries of the social network.

% Hub agents maintained a neutral representational structure that oscillated between the two clusters throughout the simulation.
% Agents $k = 0, 7$, and $13$, which serve as inter-cluster bridge hubs and are directly connected to each other, exhibited oscillatory trajectories positioned between the two cluster trajectories in the MDS embedding.
% This pattern indicates that simultaneous exposure to both communities prevented these agents from aligning with either one alone.

% Point-wise representational divergence preceded structural divergence, suggesting a two-stage differentiation process.
% Differentiation in Wasserstein distance began before time step 3000, whereas divergence in GW distance did not emerge until after time step 3000.
% This temporal dissociation indicates that agents in different clusters first developed distinct per-observation representation values while still sharing an equivalent relational structure.
% Only as cluster-specific differences in created observations accumulated did the relational structure itself begin to diverge.

Hub agents exhibited a distinct trajectory that reflected their simultaneous membership in both communities.
Agents $k = 0, 7$, and $13$, which bridge the two clusters through direct inter-cluster connections, did not converge toward either cluster's trajectory in the MDS embedding (Fig.~\ref{fig:gw-mds}) but instead remained in the intermediate region, oscillating between the two groups throughout the simulation.
This behavior is consistent with the expectation that agents exposed to competing social priors from both clusters cannot settle into a single cluster-specific representational structure.

Representational differentiation proceeded in two stages: agents first diverged in the values assigned to individual observations, and only later diverged in the relational structure among those values.
The Wasserstein distance matrix shows clear block-diagonal structure by time steps 1001--2000 (Fig.~\ref{fig:wasserstein-snapshot}), indicating that agents in different clusters had already developed distinct per-observation representations.
In contrast, the MDS embedding of the GW distance matrix shows that agents from both clusters remained tightly overlapping until approximately step 2000, with cluster-level separation emerging only thereafter (Fig.~\ref{fig:gw-mds}).
This temporal lag suggests that local differences in created observations must first accumulate sufficiently before they reshape the global geometry of each agent's representational space.

\subsection{Temporal Evolution of Memorized Observations}

\begin{figure*}[t]
    \centering
    \includegraphics[width=\linewidth]{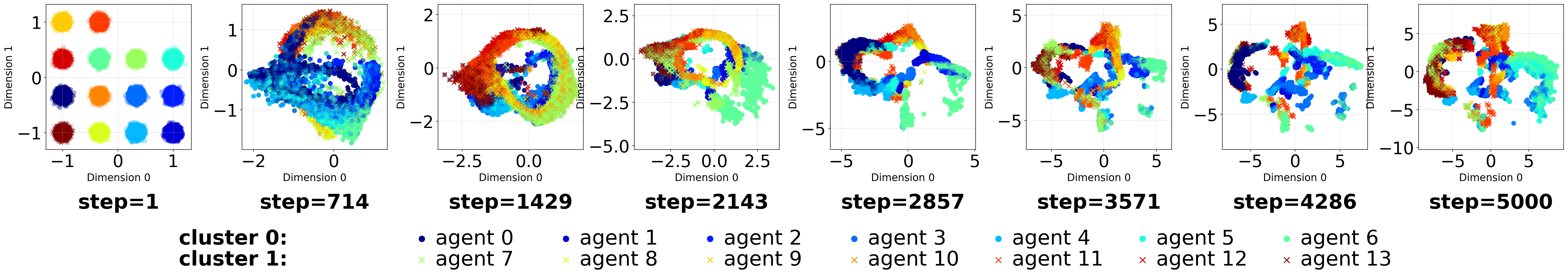}
    \caption{Temporal evolution of the distribution of observations created and memorized by each agent.}
    \label{fig:memorized-obs}
    \vspace{-4mm}
\end{figure*}

The observations created and memorized by each agent underwent a characteristic temporal trajectory: initial convergence aligned with the cluster structure, followed by the emergence of individually distinctive observations within each cluster.
Figure~\ref{fig:memorized-obs} shows the temporal change in the distribution of observations under the \texttt{w/ creation} condition.
At the start of the simulation, each agent's memorized observations were distributed across distinct positions in the observation space.
From the onset through step 1500, these observations gathered into arch-shaped structures on a cluster-by-cluster basis, with each cluster developing a characteristic distribution and internal geometry.

Following this cluster-level consolidation, individual agents began to construct genuinely novel observations beyond the existing observation region.
Notably, agents such as agent~6 and agent~13 were among the first to expand the occupied region of the observation space.
Toward the end, the coherent cluster-level structures began to dissolve, and individual agents progressively created and memorized observations unique to themselves.
At this stage, a small number of agents came to occupy nearby positions in the observation space, suggesting the emergence of finer-grained affinities that cut across the original cluster boundaries.

This temporal trajectory can be interpreted as the outcome of an adversarial interplay between social conformity, induced by model updates in response to neighbors' observations, and creative deviation, driven by each agent's generative action under active inference.
In the earlier phase, agents prioritized conforming to their immediate social context through mutual creation and memorization within clusters, gradually constructing a shared observation structure.
Once this cluster-level alignment was sufficiently established, agents shifted toward creative exploration, maintaining individual distinctiveness while remaining loosely connected to their cluster's shared structure.
This dynamic reflects how each agent actively constructed its external world under both creative curiosity and social pressure.

\subsection{Structural Similarity Between Observations and Social Representations}

\begin{figure}[t]
    \centering
    \includegraphics[width=\linewidth]{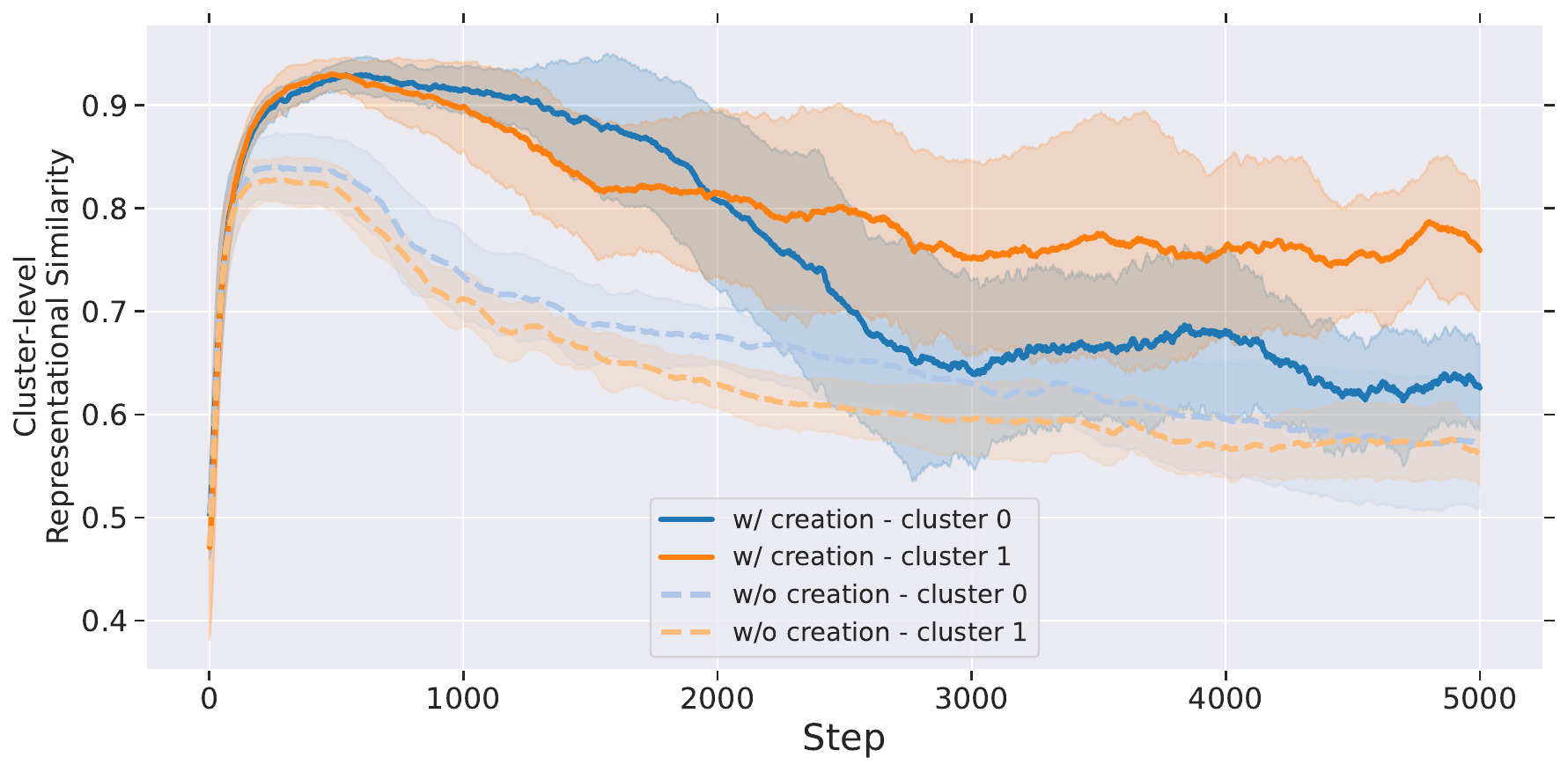}
    \caption{RSA similarity between observations and social representations for each cluster, smoothed with a 25-step moving average. Shaded bands indicate standard deviation.}
    \label{fig:cluster-rsa}
    \vspace{-4mm}
\end{figure}

To examine the effect of agents' creative acts on social representation formation, we compared the structural similarity between memorized observations and inferred social representations under the \texttt{w/ creation} and \texttt{w/o creation} conditions.
Simulations were conducted with five different random seeds for each condition.
For each cluster, we quantified the similarity using representational similarity analysis (RSA)~\citep{Kriegeskorte2008-vh} applied to the Euclidean distance matrices of the memorized observations and the social representations.
Figure~\ref{fig:cluster-rsa} shows the temporal evolution of this similarity under both conditions.

Agents' creative acts were essential for maintaining structural alignment between observations and social representations.
Under the \texttt{w/o creation} condition, similarity decreased monotonically after step 500 in both clusters.
In contrast, under the \texttt{w/ creation} condition, similarity remained at approximately 0.65 or above throughout.

This finding implies that creative acts function as a mechanism through which the structure of social representations is projected onto the observation space.
That is, the regulatory influence of social representations extended beyond the agents to shape the external world, with individual agents as the mediating mechanism.
These results indicate that the top-down regulatory influence of social representations and the bottom-up creative acts of individual agents operate in a mutually reinforcing cycle, through which social representations and the observation distribution are co-constituted.
% These results indicate that top-down regulation by social representations and bottom-up creative acts operate in a mutually reinforcing cycle, co-constituting social representations and the observation distribution over time.

\subsection{Propagation of Social Representations and Creations}

\begin{figure*}
    \centering
    \includegraphics[width=0.93\linewidth]{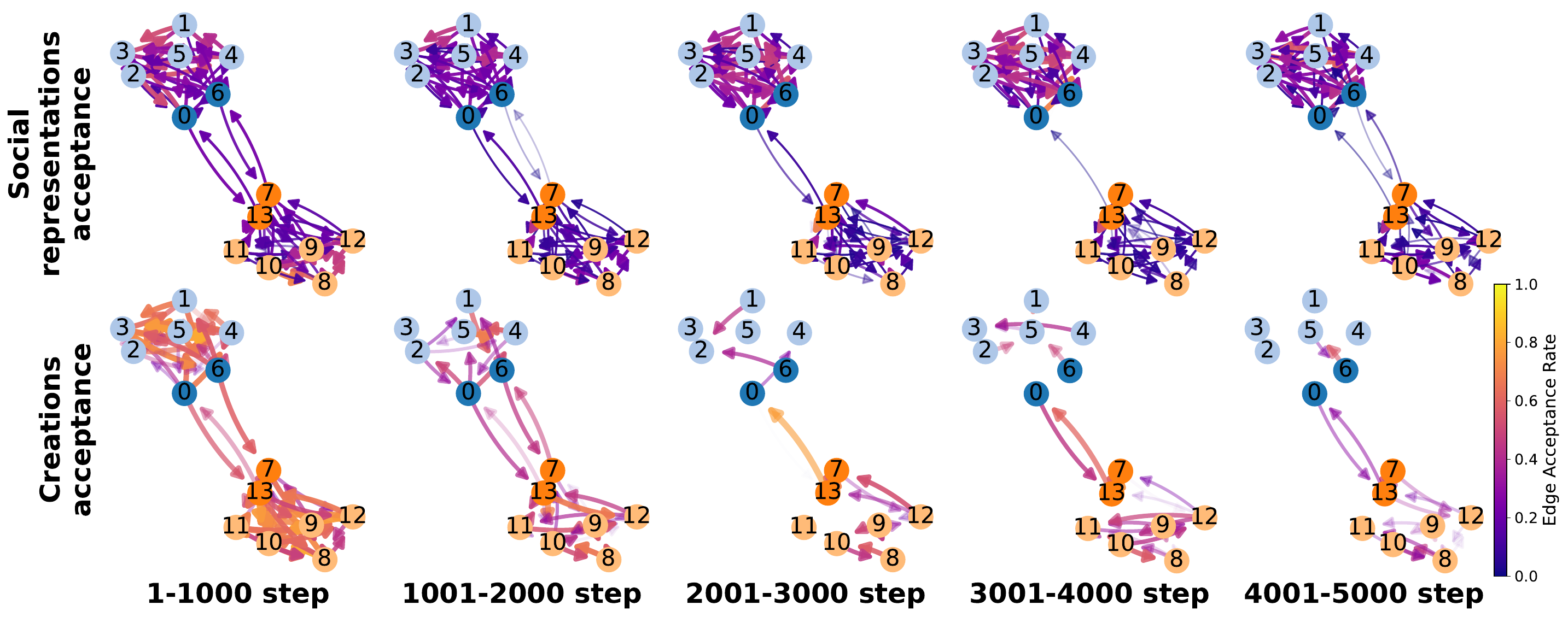}
    % \caption{Average acceptance rate of social representations and creations between agents, aggregated over successive intervals of time steps. Edge transparency reflects the frequency with which communication resulted in a zero acceptance rate.}
    \caption{Average acceptance rate of social representations and creations between agents, aggregated over successive intervals.
    Edge width and color encode the mean acceptance rate; transparency indicates the proportion of zero-acceptance time steps.}
            \label{fig:accept_network}
    \vspace{-4mm}
\end{figure*}

We analyzed the acceptance rates of social representations and creations across agent pairs to investigate their propagation dynamics (Fig.~\ref{fig:accept_network}).
Edge transparency reflects the frequency with which communication resulted in a zero acceptance rate; edges become more transparent as the proportion of zero-acceptance time steps increases.

The stable intra-cluster propagation of social representations accounts for the sustained representational alignment observed across the preceding analyses (Figs.~\ref{fig:wasserstein-snapshot} and \ref{fig:gw-mds}).
Within each cluster, edges were rarely pruned, and social representations were consistently accepted in a reciprocal manner throughout the simulation.
This persistent mutual acceptance generated continuous pressure for each agent to internalize fellow cluster members' representations, anchoring individual generative models to the shared cluster structure.
Meanwhile, inter-cluster edges underwent pruning in the latter half, indicating that social representations became increasingly difficult to share across cluster boundaries.

In contrast, the propagation pathways of creations shifted dynamically over time, accounting for the transient inter-agent similarities observed in the observation space (Fig.~\ref{fig:memorized-obs}).
Within clusters, edge pruning occurred such that creations came to be shared only among a small subset of agent pairs, and these pruning patterns shifted over a large timescale.
Across clusters, creations were consistently accepted between the hub agents, agent~0 and agent~13.
Furthermore, in the latter half of the simulation, agent~0 frequently shared creations with agents in cluster~1 while ceasing to share with agents in its own cluster~0.
The observation that a small number of agents came to produce similar creations at any given time can thus be attributed to these dynamically reconfiguring acceptance pathways.
The temporal shift in such agent combinations further reflects the ongoing reconfiguration of acceptance patterns.
For instance, the sustained convergence between agent~0 and agent~13 corresponds to their persistent mutual acceptance across cluster boundaries.

\section{Discussion}

% This study presents a computational model of social reality construction that integrates the regulatory influence of social structure with individual creative action under the framework of active inference.
% Agents are formalized as generative models that are continuously updated through communication on a social network, capturing both the internalization of social norms and the generation of novel observations through creative action.
% This dual formulation allowed us to investigate the mechanisms by which social representations emerge, stabilize, and diverge over time.

The simulation results demonstrate that the model gives rise to informationally coherent social groups whose representational structure mirrors the cluster topology of the underlying network.
Agents within the same cluster converged toward similar generative models, while inter-cluster divergence became progressively more pronounced.
Moreover, in the \texttt{w/ creation} condition, a \emph{circular mutual constitution} emerged: agents' generative models shaped the direction of their creative outputs, which propagated through the network and in turn reshaped the social representations of other agents.
The persistent structural alignment between representational and observational spaces confirms that creative action is functionally integral to the coherence of social reality, not merely a byproduct of individual expression.

The analysis of propagation dynamics further reveals a dual mechanism that sustains social representations amid continuously changing observations.
Within each cluster, sustained mutual acceptance of representations through the naming game generates a persistent internalization pressure that draws individual models toward a shared attractor.
This pressure maintains collective coherence even as the observational landscape is continually altered by creative action.
In contrast, the reception of creative artifacts became progressively concentrated among specific agent pairs, indicating that agents effectively constructed cultural niches.
% ---shared observational spaces meaningful within particular relational contexts but not others.
This selective uptake points to a process of cultural differentiation driven not by top-down prescription but by the self-organizing dynamics of local interaction.

The observed conformity arises not from social norms acting as an independent causal force, but as an emergent consequence of the complementary interplay between model update and creative action.
Model update draws each agent's generative model toward the social prior constructed through local interaction, while creative action drives exploration of observations that diverge from this prior without disrupting epistemic homeostasis within its generative model.
% The conformity so achieved in turn provides a stable basis from which creative exploration can proceed.
The conformity so achieved, in turn, provides a stable basis for creative exploration.
Crucially, no variable encodes a directive to conform, and the social prior is not an external shared variable but is constructed internally by each agent through the naming game.
Although each agent actively seeks to diverge, the generative model from which it departs is continually drawn back toward the social prior, indirectly confining the scope of deviation.
What an external observer would identify as a social norm thus arises as an epiphenomenon of this cycle, yet it exerts genuine regulatory influence, because the collective pattern feeds back into each agent's social prior and constrains subsequent cognition and action.

The present model has several limitations that suggest directions for future work.
First, the social network is fixed throughout the simulation, precluding active partner selection and the self-organized emergence of communication topology.
Because creative exploration is jointly constrained by each agent's generative model and this static structure, the model also cannot capture paradigmatic shifts in which the representational framework itself is restructured.
Treating spatial position and movement as variables subject to active inference would let the topology co-evolve with agents' generative models, so that social relationships develop alongside representational change, thereby potentially enabling such transitions.
Second, the model abstracts away embodiment, such that agents generate observations without mediation by a body or physical environment.
Such an environment would serve not only as a medium for creative acts but also as a material substrate that constrains and channels their outcomes.
Incorporating action variables operating through a body--environment system would capture the full causal chain from creative intention to observable artifact.
Third, the study relies on fixed hyperparameters, and a systematic comparison across parameter settings would clarify how each shapes the dynamics of social reality construction.
Beyond these extensions, information-theoretic tools such as non-trivial information closure (NTIC) \citep{Chang2020-ny, Dobata2025-rh} offer a complementary analytical direction.
Applying NTIC to quantify the degree of autonomy in each agent's behavior would provide a formal measure of social autonomy, contributing to a mathematical understanding of the emergence of social self in Mead's sense.

\section{Conclusion}

% This study presents a computational model of social reality construction that integrates the regulatory influence of social structure with individual creative action under the framework of active inference.
% Agents are formalized as generative models that are continuously updated through communication on a social network, capturing both the internalization of social norms and the generation of novel observations through creative action.
% This dual formulation allowed us to investigate the mechanisms by which social representations emerge, stabilize, and diverge over time.
This study presents a computational model of social reality construction that integrates the regulatory influence of social structure with individual creative action under active inference.
Agents are formalized as generative models continuously updated through communication on a social network, capturing both the internalization of social norms and the generation of novel artifacts through creative action.
This dual formulation lets us investigate how social representations emerge, stabilize, and diverge over time.

\section{Acknowledgements}
This work was supported by JSPS grant JP23H04834.
Generative AI tools (Claude, Anthropic) were used to assist with code development and manuscript preparation.

\footnotesize
\bibliographystyle{apalike}
\bibliography{src/reference}

\end{document}